\documentclass[conference]{IEEEtran}
\IEEEoverridecommandlockouts
\usepackage{cite}
\usepackage{amsmath,amssymb,amsfonts}
\usepackage{algorithmic}
\usepackage{graphicx}
\usepackage{textcomp}
\usepackage{xcolor}
\usepackage{acro}
\def\BibTeX{{\rm B\kern-.05em{\sc i\kern-.025em b}\kern-.08em
    T\kern-.1667em\lower.7ex\hbox{E}\kern-.125emX}}

\DeclareAcronym{AAS}{
short = AAS,
long = {Asset Administration Shell}
}
\DeclareAcronym{SM}{
short = SM,
long = submodel,
long-plural = s
}
\DeclareAcronym{C4I}{
short = C4I,
long = Capability for Industry
}
\DeclareAcronym{OWL}{
short = OWL,
long = Web Ontology Language
}
\DeclareAcronym{SME}{
short = SME,
long = {submodel element},
long-plural = s
}
\DeclareAcronym{SMC}{
short = SMC,
long = {submodel element collection},
long-plural = s
}
\DeclareAcronym{Id}{
short = Id,
long = identification,
}
\DeclareAcronym{PPR}{
short = PPR,
long = {product, process and ressource},
}
\DeclareAcronym{MTP}{
short = MTP,
long = {Module Type Package},
long-plural = s
}
\DeclareAcronym{PEA}{
short = PEA,
long = {Process Equipment Assembly},
long-plural-form = {Process Equipment Assemblies}
}
    
\begin{document}

\title{An Ontology-Based Capability Description of MTP Modules}

\author{\IEEEauthorblockN{Christian 
Barth\IEEEauthorrefmark{1}, Tobias Klausmann\IEEEauthorrefmark{2}, Andreas Bayha\IEEEauthorrefmark{3}, Matthias Freund\IEEEauthorrefmark{1}, Kathrin Gerber\IEEEauthorrefmark{4}} 
\IEEEauthorblockA{\IEEEauthorrefmark{1}Festo SE \& Co. KG,
Esslingen, Germany,
\{christian.barth, matthias.freund\}@festo.com}
\IEEEauthorblockA{\IEEEauthorrefmark{2}Lenze SE,
Aerzen, Germany,
tobias.klausmann@lenze.com}
\IEEEauthorblockA{\IEEEauthorrefmark{3}fortiss GmbH,
Munich, Germany,
bayha@fortiss.org}
\IEEEauthorblockA{\IEEEauthorrefmark{4}Festo Didactic SE,
Denkendorf, Germany,
kathrin.gerber@festo.com}
}

\maketitle

\begin{abstract}
The \ac{MTP} is an emerging standard for the software integration of process modules into a control system. The core concept of \ac{MTP} is to separate process plants into autonomous modules called \acp{PEA} which offer easy to use high level services. Applying the \ac{MTP} concept leads to a modular production which can easily be adapted by adding, subtracting, or exchanging individual units. In order to leverage this flexibility, it is important to know the capabilities each \ac{PEA} offers and expose them in a machine-readable way via its digital twin. As one realization of this requirement, this paper describes how a semantic annotation of \ac{MTP} services using an ontology can be realised in conjunction with the \ac{AAS} technology.   
\end{abstract}


\section{Introduction}

One of the key aspects of modern automation is the decentralization of automation logic with a service-based architecture. Individual components or modules consisting of multiple components offer easy to use and easy to integrate functionalities. 
The big advantage of this architecture is the possibility to adapt easily to changing requirements in the production process. The engineering effort to technically integrate the control logic of modules and components is very small due to standardized interfaces like \emph{\ac{MTP}} \cite{vdi2019mtp1} or \emph{PackML} \cite{packml}. Less optimized aspects of engineering are planning and purchasing. There have been first activities to systematically describe the capabilities of production modules \cite{vdi2022modularplants}. In our paper, we build upon this, aligning with concepts of specifying and classifying the capabilities semantically as proposed in \cite{weser2020ontology}. The goal is to create a formalism with which a flexible adaptation of production processes at minimal effort for most involved engineering domains is possible. 
With this approach the planning for example could be improved in the following way:
During the conception phase of a plant, an abstract process is defined. This is used to derive a set of capability requirements needed to execute the planned actions. Based on the capability descriptions of modules proposed in this paper, it is then possible to compile lists of suitable candidates automatically, reducing significantly the effort to select appropriate modules. 

The remainder of the paper is structured as follows:
Section \ref{sota} gives an overview of the most relevant technologies for the paper topic. 
We start with a brief description of the \acl{MTP} including its purpose and current state. This is followed by a summary of the \acl{AAS}, while mentioning how \ac{AAS} can complement the \ac{MTP} technology. The third and most comprehensive subsection is about Capabilities \& Skills. Besides a general description of the concept, we explain the use of ontologies to describe capabilities and their relations. Furthermore we introduce the \emph{capability submodel} for \ac{AAS}.
In Section \ref{approach}, we introduce our proposal for a capability ontology for \acp{PEA} together with implications for the \emph{capability submodel}, allowing for (semi-) automatic selection of modules. 
In Section \ref{case_study}, we describe and discuss the implementation of the approach with the example of a painting module, including a critical reflection of the limitations of the approach.
The last section gives a summary of the paper and an outlook of possible future extensions.

\section{State of the Art}
\label{sota}
\subsection{\acl{MTP}}\label{MTP}
Modular production is an approach to tackle the growing demand for shorter product lifespans, smaller lot sizes, and increased flexibility. To be able to implement this approach in the context of the process industry, several working groups of NAMUR (German user association of automation technology in process industries) and ZVEI (German association of automation technology vendors) are currently developing the \emph{\acf{MTP}} \cite{vdi2019mtp1}. \ac{MTP} thereby unifies the (automation) interface of a process module to enable its integration into any plant without the need for complex integration processes. This shall enable faster engineering \& reconfiguration times for process plants as they can rely on a set of pre-engineered process modules.

\ac{MTP} files are based on AutomationML \cite{iec2018aml} and contain relevant information about a module type: A description of the provided communication and control interfaces of the components \cite{vdi2020mtp3} (e.g. on an OPC UA server), an abstract representation of an HMI \cite{vdi2019mtp2}, and a description of services (e.g. dosing or stirring) offered by the module \cite{vdi2020mtp4}. The general structure of these services is standardized among others by means of a common state machine that each service must implement. However, \ac{MTP} does not specify a common set of standard service functionalities but leaves the service design and naming to the module vendors. This makes finding the optimal module and service for a certain process step a complex task for the plant engineer.

Important relationships between the domains of process engineering and (\ac{MTP} based) automation of a process plant are discussed in \cite{vdi2022modularplants}: Each \ac{MTP} service realizes one or multiple \emph{process functions}\footnote{Translated from German: "Prozesstechnische Funktion"} that can be separated into main functions and utility functions. While \cite{vdi2022modularplants} gives examples for such functions (e.g. dosing, tempering, or pressure adjustment), no complete list is given. Each of these functions can also be characterized by different attributes that again are only listed exemplary (e.g. maximum throughput for dosing). The guideline also discusses the description of services offered by a module and/or of service requirements for a modular plant as well as a procedure for the manual or semi-automatic selection of process modules' services based on these \emph{process functions}. However, \cite{vdi2022modularplants} does not state how services can be described or annotated in a machine-readable way, which is necessary for the (semi-)automatic selection of modules' services. 

\subsection{\acl{AAS}}
\label{aas_metamodel}
\begin{figure}[htbp]
    \centering
    \includegraphics[width=\linewidth]{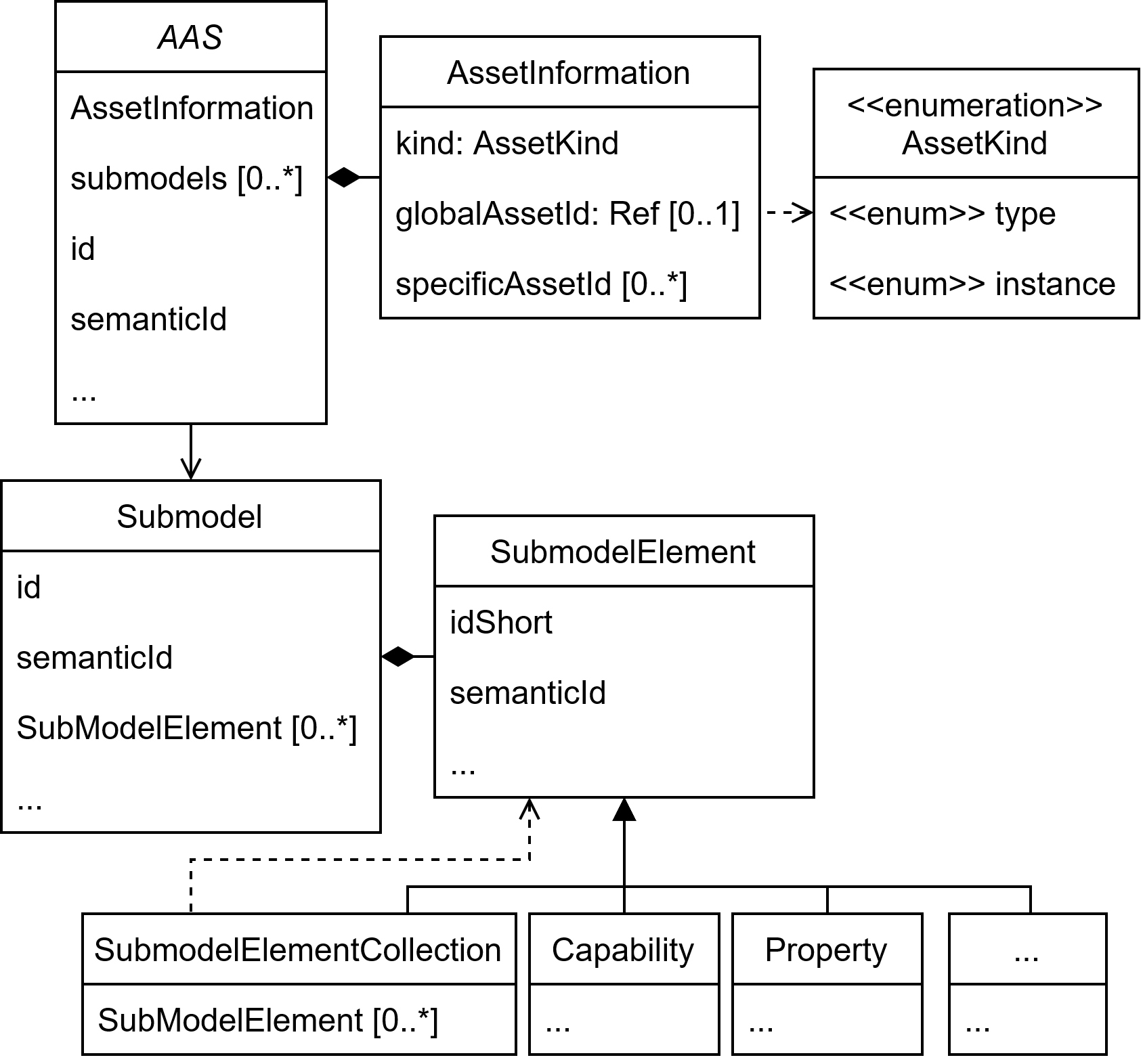}
    \caption{
    Overview of a selection of \ac{AAS} meta model elements
    }
    \label{fig:aas_overview}
\end{figure}

To exchange information for components and modules, a digital twin \cite{PlattI4_0:Glossary:D} is a useful  container of the necessary information. As one implementation of a digital twin, the \ac{AAS} meta model \cite{aasp1v3rc1} can be used. The meta model consists of several sub-elements. A selection is presented here to reflect the basic structure of an \ac{AAS}:
\begin{itemize}
    \item The \acl{AAS} entry element serves as a container with references to the \aclp{SM} of the \ac{AAS} as well as an asset information sub element.
    \item The asset information element contains one global unique asset \ac{Id} and may contain several specific asset \ac{Id}s which are only unique in a local specific scope of a plant or product line etc. Furthermore, the asset kind can be distinguished between type and instance assets, reflecting e.g. a product line or a single particular product.
    \item The \acl{SM}s contain a varying number of \acp{SME}.
    \item The \acl{SME}s express the information of an \ac{AAS}, e.g. a textual string or a number. The \ac{SME}s can be nested in \acp{SMC}.
\end{itemize}
An overview of the \ac{AAS} metamodel structure is presented in Figure \ref{fig:aas_overview}.
All \ac{AAS} elements can be semantically annotated by using their attribute "semantic\ac{Id}". The \ac{AAS} submodels are additionally identified by a globally unique \ac{Id} and thus can be referenced in arbitrary information models. The \ac{SME}s can be referenced using the \ac{SM} \ac{Id} and a \ac{SME} idShort path list inside the corresponding \ac{SM}. 

In the scope of the BaSys 4.2 project\footnote{https://www.basys40.de/}, a proposal has been made to represent modules specified by the MTP standard within \emph{Asset Administration Shells} (AAS) \cite{gruener2021mtpaas}. The concept is currently being specified as an AAS submodel in the Industrial Digital Twin Association (IDTA)\footnote{https://industrialdigitaltwin.org/}. The IDTA is the host organization for AAS submodels as industry standards. 
Even though \ac{MTP} explicitly focuses on the engineering phase of module types (i.e., aspects regarding the distribution of MTP files or the operation of a module are not discussed), there are two types of AAS models specified: an AAS for a type asset in which the MTP description file is attached or linked, and an \ac{AAS} of an instance asset providing \ac{PEA}-specific information which is relevant for actual operation of the module. These are for example the OPC UA endpoints and an identifier for the actual position and role of the module. The main benefit of having the MTP aspects represented in an AAS is to allow for relations and links between entities described in the MTP manifest and information modeled in other AAS submodels. These could e.g. be documentation or bill of material.

\subsection{Capabilities \& Skills}
\label{cap_and_skills}
To achieve the previously mentioned flexibility improvements of automation, a systematically modelled description of the various parts of a production system is required.
The \ac{PPR} model \cite{PPR} describes the aspects of production as a relationship between the products (e.g. work piece), the production process (e.g. recipes) and production resources (e.g. automation equipment). The \emph{capabilities} and \emph{skills} influence all three aspects and act as a connecting concept. In this paper we use the terms for \emph{capability} and \emph{skill} as defined in \cite{weser2020ontology}: "\emph{Skills} represent the parameterizable and executable functionality
of hardware or software components, and \emph{capabilities} provide a meta level description of their effects." Thereby, "meta level" refers to product, process and resource independence. 

\begin{figure*}[!htbp]
    \centering
    \includegraphics[width=0.8\linewidth]{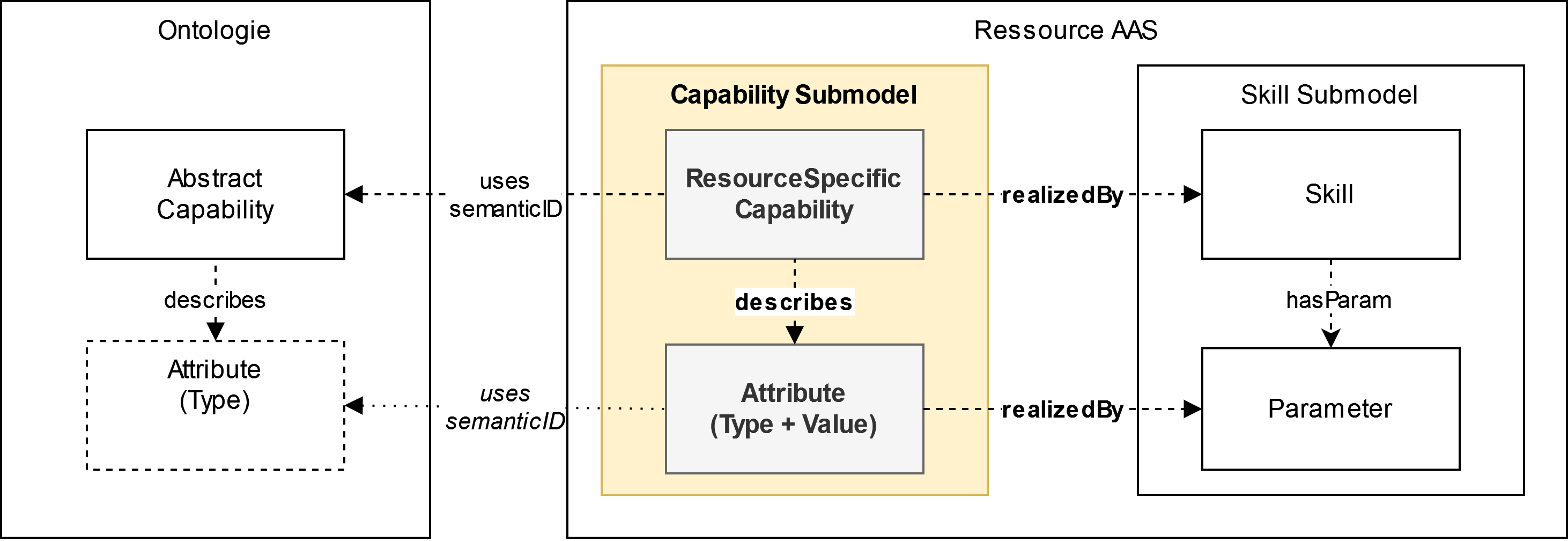}
    \caption{
    The capability \ac{SM} in relation to ontologies and skills
    }
    \label{fig:cap_sm_refs}
\end{figure*}

In this paper, we distinguish between \emph{abstract capabilities} and \emph{resource specific capabilities}, refining the original \emph{capability} term: An \emph{abstract capability} provides a meta level description of an effect along with meta level properties for a detailed description, which can be defined in \emph{capability ontologies} (see section \ref{cap_ontologies}). A \emph{resource specific capability} \cite{jwp_cap_skill, aas2020details} is based on the \emph{abstract capability} and is only valid in the context of a resource. With the specified context, the meta level elements can be instantiated as model elements, for example by using the \ac{AAS} meta model elements, see Section \ref{aas_metamodel}. The model elements can now be filled with information reflecting the context of the resource and the effect it can have in a production context.

Capabilities have a describing nature and can not cause the desired effect in the real world. To achieve this, a \emph{skill} is used, which is the executable counterpart of a \emph{capability}. There are several options available to implement a \emph{skill} as described in \cite[Section~5.4]{jwp_cap_skill}. 
A skill is based on the information provided by a \emph{resource specific capability}, potentially extending the level of detail of the model with e.g. environmental conditions or implementation specifics. With the \emph{resource specific capabilities} and the \emph{skills} available, a production process, which brings certain requirements to manufacture a product, can use the \emph{capabilities} and the \emph{skills} as follows: Based on the requirements, the resources can be queried in a two step procedure. First, a shallow match can be executed using the requirements and the \emph{resource specific capabilities}, named \emph{capability matching}. Following a successful \emph{capability matching} step, a more detailed \emph{skill matching} that takes the additional skill description into account can be conducted to assure that all requirements for the execution of process are met, see \cite{functionalCapabilitySkill}. The two checks are described in more detail in \cite[Chapter~6]{jwp_cap_skill}.


\subsubsection{Capability Ontologies}
\label{cap_ontologies}

An ontology is a formal representation of data and relations. It uses logically specified edges to model knowledge in a way that implicit relationships can be inferred by logic interpreters or reasoners. In this way, modelled information can be interpreted automatically and different sources of information can be combined effectively. The ontology is a semantic technology which offers high flexibility in modeling concepts. This leads to the problem of finding a suitable modeling structure. In \cite{weser2020ontology}, a well-established approach has been chosen, which consists in dividing the modeled concepts in terms of general understanding, specific taxonomies, and actual data into so-called upper ontologies, domain ontologies, and application ontologies, respectively. Since the \ac{OWL}\footnote{https://www.w3.org/OWL/} standard was chosen here, its import declaration can be used to represent dependencies between these ontologies. This \ac{C4I}\footnote{downloadable at: https://www.w3id.org/basyx/c4i} ontology is presented in \cite{weser2020ontology}.

\subsubsection{\ac{AAS} Capability Submodel}
\label{cap_sm}
While ontologies are an adequate concept to model the abstract capabilities themselves, in order to identify capabilities of a given resource, these ontological descriptions need to be linked with the assets providing the actual skills and implementing these capabilities. 
As described in Section \ref{aas_metamodel}, when using the \ac{AAS} for the asset's description, \ac{AAS} \acp{SM} are used to model the various aspects of production resources. 
The capability submodel proposed in~\cite{jwp_cap_skill} is dedicated for describing the capabilities provided by an asset.
 
\begin{figure*}[!ht]
    \centering
    \includegraphics[width=0.7\linewidth]{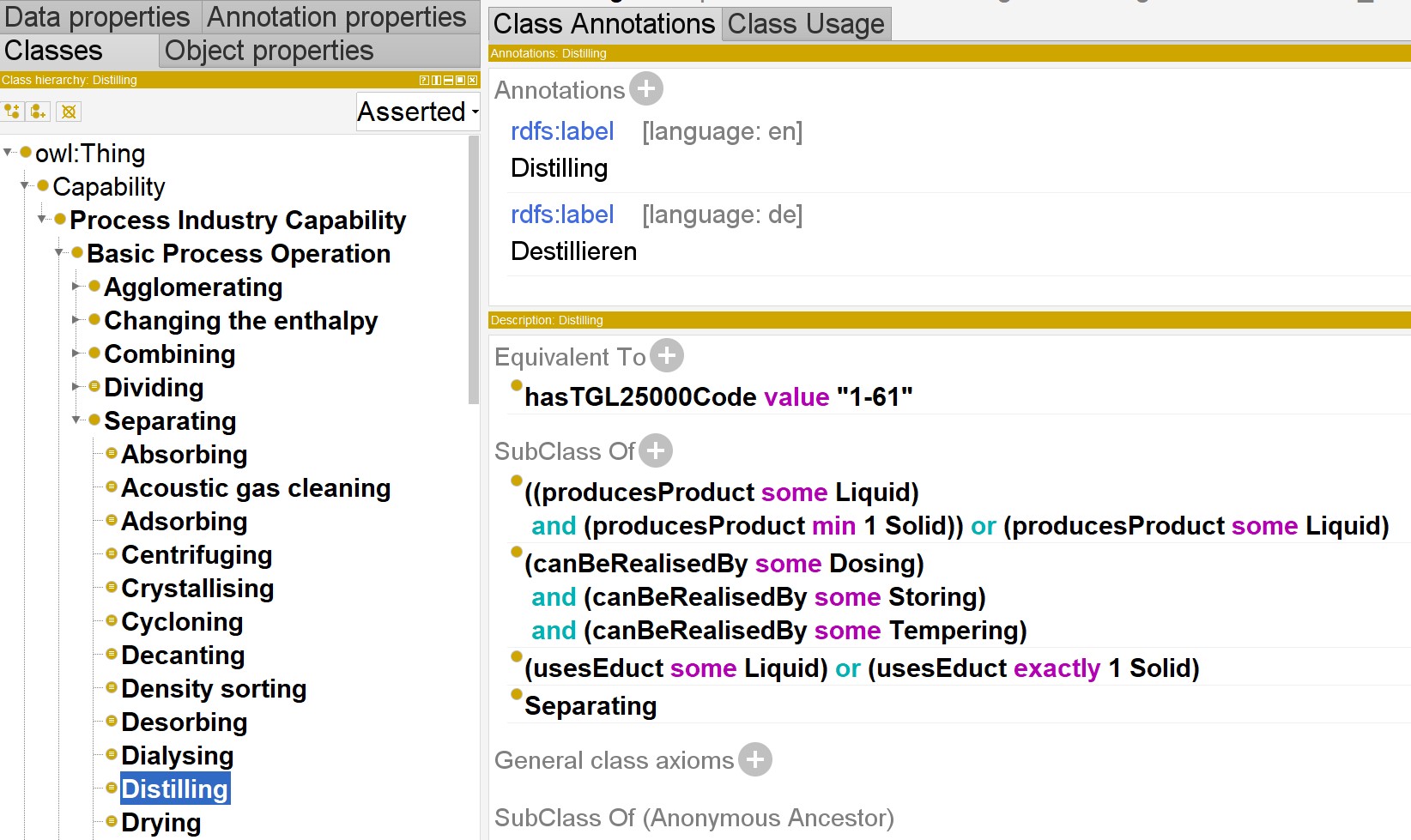}
    \caption{
    Capability \emph{distilling} in process automation domain ontology based on TGL 25000 standard.
    }
    \label{fig:distilling}
\end{figure*}

Figure~\ref{fig:cap_sm_refs} gives an overview about these notions and their relations with each other: 
\emph{ontologies} model the abstract capabilities themselves, while the \ac{AAS} of the resources contain a \emph{capability submodel} that describes the provided capabilities of this specific resource or resource type. 
Within the capability \ac{SM}, the dedicated \emph{capability} model element of the \ac{AAS} meta model can be used to model the \emph{resource specific capabilities}.
Hereby, the \emph{semanticIDs} of these capability elements reference the \emph{abstract capabilities} in the ontology and associate them with the \emph{resource specific capabilities} of the asset.
Besides this reference to the abstract capabilities, the capability \ac{SM} also contains the reference to the executable implementation of the capabilities, i.e. to the \emph{skills}. 
In accordance with Fig.~\ref{fig:cap_sm_refs} this reference is named \emph{realizedBy}.

A similar reference as between the capabilities and the skills can be seen between the capabilities' \emph{attributes} and the skill's \emph{parameters}:
while abstract capabilities in ontologies describe which attributes are relevant for an abstract capability, the resource specific capabilities describe the actual values of the resource's capabilities. 
Finally, the skill which realizes the resource specific capability could determine this value for every skill execution via a \emph{parameter} that is taken into account for the skills execution.

An example for these notions could be the abstract capability \emph{dosing} that can be further described with an attribute \emph{amount} in an ontology. 
A concrete resource which implements \emph{dosing} could then offer a resource specific capability \emph{dosing} with a semantic ID to this ontology and an attribute \emph{amount} with a value range of \emph{0.2ml to 80.0ml}.
Finally, the skill which implements \emph{dosing} can have a parameter \emph{amount} that can be set to the value \emph{65ml} to make the skill dose this amount of liquid in it's next execution.

With these three concepts of abstract capabilities, resource specific capabilities, skills and the references between them, one can seamlessly find related information:
for every skill it is possible to see which capabilities it realizes; for every abstract capability it becomes possible to find resources that provide this capability by the skills in their implementation.

\section{Approach}
\label{approach}

As described in Section~\ref{cap_and_skills}, a skill represents the \emph{parameterizable and executable functionality of hardware or software}. Regarding the context of \ac{MTP}, this is implemented by the concept of \emph{MTP services} (cf.  Section~\ref{MTP}): While the implementation of a service may be hidden by the module vendor, services can be directly invoked on a module via a set of variables provided on a public OPC UA server. Furthermore, services can also be parametererized\,--\,the set of parameters for each service is thereby defined in the \ac{MTP} file.

Based on this analogy and the integration of \ac{MTP} into the \ac{AAS} described in Section~\ref{aas_metamodel}, the \ac{MTP} concept can be combined with the concept of capabilities described in Section~\ref{cap_and_skills}. The motivation of separating the capability information into a representation within \ac{AAS} on the one side and the ontology on the other side is the following: The ontology can be used as an industry wide single source of truth, providing generally valid relations between \emph{process functions}, \emph{basic process operations}, educts and products. By limiting the capability modelling in the ontology to these building blocks, one avoids high complexity and modifications making it easy to integrate it e.g. into an engineering tool. With the help of reasoners translations between the the different elements of the above-mentioned ontology content is possible. The \ac{AAS} representation on the other hand is meant to hold implementation specific information, which can vary significantly from one application to another. By describing the combination of \emph{process functions} used to implement the specific PEA service functionalities and associating them to the respective ontology elements all relevant information for selecting modules according to different requirement specification is accessible. The downside of the approach is that the tooling needs to support both \ac{AAS} and ontology technologies. Since the \ac{AAS} ecosystem offers solutions for various engineering use cases, the additional effort of implementing \ac{AAS} support can be used in multiple applications.    

The approach presented above can be used to enrich the \ac{MTP} ecosystem with the currently missing aspects of automatically finding/selecting \acp{PEA}/services (cf. Section~\ref{MTP}) that can be realized based on the capability approach. 
Hence, this section first presents an extension of the C4I ontology (cf. Section~\ref{cap_ontologies}) that covers specific aspects of the process industries. After that, resulting implications on the capability submodel (cf. Section~\ref{cap_tm_implications}) are discussed. 

\subsection{Ontology for MTP Modules}

As already discussed in Section \ref{cap_ontologies}, a meta model for a capability ontology was developed in \cite{weser2020ontology}. For factory automation the authors propose a handling domain ontology on the basis of the VDI guideline 2860. 
However, for process automation no such domain ontologies exist so far. As was also described in \cite{weser2020ontology}, it is reasonable to take existing state of the art into account. For this reason, we searched for an existing taxonomy for process automation tasks and found a suitable candidate in the TGL 25000 standard \cite{tgl25000}. This standard offers a categorization of so called \emph{basic process operations}\footnote{English translation from the German term \emph{Verfahrenstechnische Grundoperationen}}. It also categorizes them based on three states of matter: solid, liquid and gas. Additionally it defines educt and product per operation following these three types. We created an process industries specific domain ontology based on the informations listed in the TGL standard \cite{tgl25000}. To project the educts and products in the ontology, we introduced the object properties \emph{usesEduct} and \emph{producesProduct}.

The example of \emph{distilling} as depicted in Figure \ref{fig:distilling} demonstrates how these concepts were realized in the ontology: The basic process operations of TGL 25000 \cite{tgl25000} were modeled as classes under a class \emph{Basic Process Operation} under the main class \emph{Capability} as proposed in \cite{weser2020ontology}. Additionally, we modelled the hierarchical categorization as a class hierarchy, as it can be seen in the example of \emph{distilling} in Figure \ref{fig:distilling}: \emph{Distilling} is a subclass of \emph{separating}. In addition to the hierarchical categorization we modelled a rule-based sub classification, further specifying the underlying principles of the capability as can be seen in the following: \emph{Distilling}
can also be described in terms of products it can produce or educts it uses. For both, in this example there are two possibilities as described in TGL 25000 \cite{tgl25000}. \emph{Distilling} can either produce a non-defined amount ("some") of liquid products and at least one solid product or it can produce only a non-defined amount of liquid products. For this it uses either some liquid educts or at least one solid educt.

\begin{figure}[htb]
    \centering
    \includegraphics[width=0.5\linewidth]{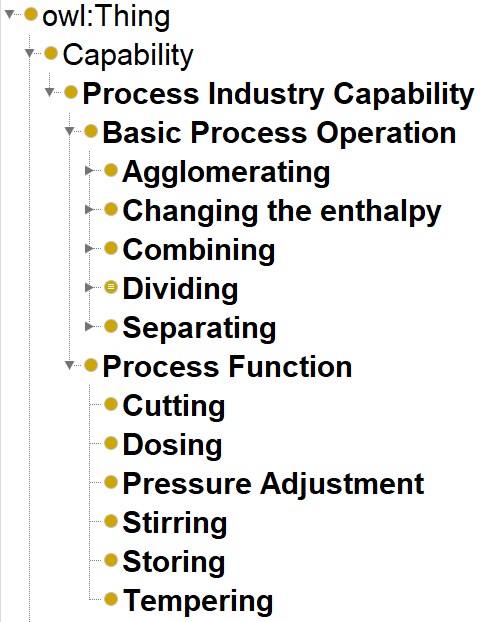}
    \caption{
    Overview over \emph{basic process operations} and \emph{process functions} in process automation domain ontology}
    \label{fig:operationsFunctions}
\end{figure}

When designing a chemical process, the \emph{basic process operations} define the modifications of the material and are thus key elements for the process technician. However, the technical realizations of these basic process operations are of importance as well as they heavily influence the application area and performance of a solution. Hence, describing the capabilities should always relate to both the \emph{basic process operation(s)} a PEA can provide as well as the technical realization that is used to realize these operations. As already introduced in Section \ref{MTP}, the technical realization can be described based on \emph{process functions}. To reflect this in the ontology, we introduced the concept of \emph{process functions} as specialized category of \emph{capability}. Figure \ref{fig:operationsFunctions} shows the resulting main structure of the ontology and the defined process functions. As there is no standard listing all or at least a decent amount of process functions that can be used in the process industry (cf. Section \ref{MTP}), the process functions that have been added to the ontology can only be seen as exemplary until a standardization is done in this regard. 

Finally, the object property \emph{canBeRealisedBy} was added to the ontology to describe possible realizations (i.e. process functions) for a basic process operation. Figure \ref{fig:distilling} shows this for the example of the basic process operation \emph{distilling} that can be realized by a combination of the three process functions \emph{dosing}, \emph{storing} and \emph{tempering}.  

\subsection{Implications for the Capability Submodel}
\label{cap_tm_implications}
As presented in the last section, basic operation capabilities are defined in an process industries specific ontology. Taking the capability \ac{SM} shown in Section \ref{cap_sm} as a basis, additional implications coming from the \ac{MTP} structure have to be addressed: An \ac{MTP} module consists of several components. Hence, the module may have to form a \emph{complex capability} that reflects its function as a composition of the built-in components' capabilities.

As an example, a \emph{painting module} may consist of a tank, a pump, a valve, and a spray nozzle. While these components on their own have the capabilities \emph{store}, \emph{pressure adjust}, \emph{dose}, and \emph{spray}, they together form the complex capability \emph{paint} which is also implemented by a specific skill. While ontologies are introduced in Section \ref{cap_ontologies} as a base for the definition of resource-neutral capabilities, a \emph{paint} capability could also be realized by different resources and their related capabilities. For example, painting could also be realized by lowering the product into an immersion bath.

To facilitate the manual or automatic selection of modules' services, both the complex capabilities and the specific underlying  capabilities shall be exposed in the capability \ac{SM} of the \ac{AAS}. Furthermore, the composition relations between the capabilities are modeled as well.
This approach can be applied for complex capabilities in general. For example the capability \emph{pick\&place} used in factory automation is composed of \emph{move} and \emph{grasp}, where \emph{grasp} may be a composition of \emph{hold} and \emph{release}.

\section{Case Study Painting Module}\label{case_study}

\subsection{BaSys Demonstrator}
\label{demonstrator}

\begin{figure}[!htbp]
    \centering
    \includegraphics[width=\linewidth]{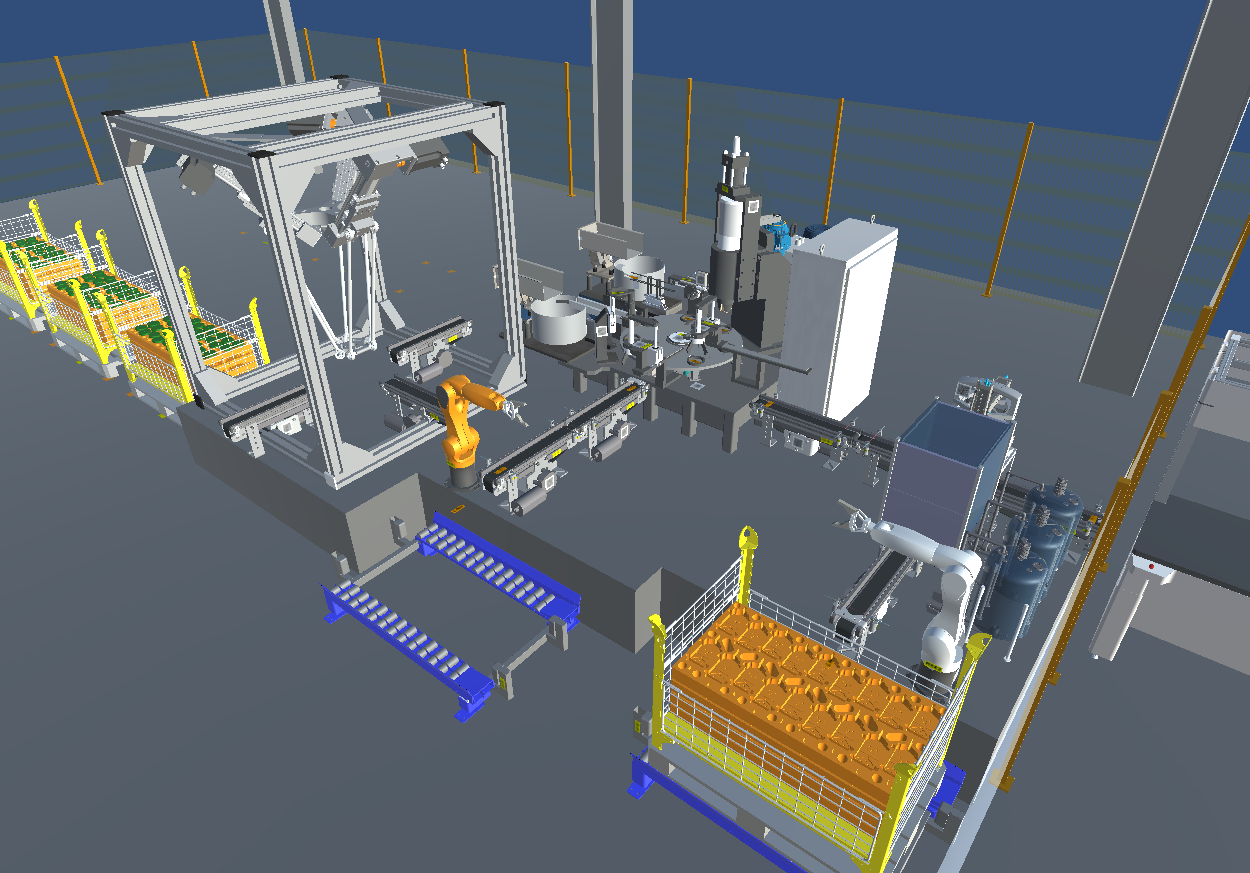}
    \caption{
    Screenshot of the visualization of the virtual BaSys demonstrator of a tin assembly and painting process.
    }
    \label{fig:basysdemo}
\end{figure}

The mostly virtual BaSys demonstrator\footnote{https://github.com/eclipse-basyx/basyx-demonstrators} highlights concepts developed in the BaSys 4.0 and BaSys 4.2 projects. A downloadable version can be used to evaluate the different aspects of an \ac{AAS}-centric production including the BaSyx middleware\footnote{https://www.eclipse.org/basyx/}.

As part of the demonstrator, a fictional process of a tin assembly and painting is implemented. It is visualized using software from \emph{unity technologies}\footnote{https://unity.com}. Fig.~\ref{fig:basysdemo} shows the layout of the virtual demonstrator. All resources that provide the automation functionalities used in the demonstrator have \ac{AAS} representations with various \aclp{SM} describing technical data, capabilities, documentation, and control component interfaces (the latter being a BaSyx specific concept of wrapping skills of components and modules and offering unified execution interfaces and state information). The control component features adapters to translate commands and states to MTP or PackML and allows thus simultaneous orchestration of traditional integration methods from process industries and factory automation. The orchestration itself is implemented using \emph{Business Process Model and Notation} (BPMN) scripts for the process descriptions. These \emph{workflows} are then evaluated and executed by the open source engine Camunda\footnote{https://camunda.com}, starting and stopping the skills of the various components and controlling the respective states. 


\subsection{Painting-Module}
\begin{figure}[htbp]
    \centering
    \includegraphics[width=\linewidth]{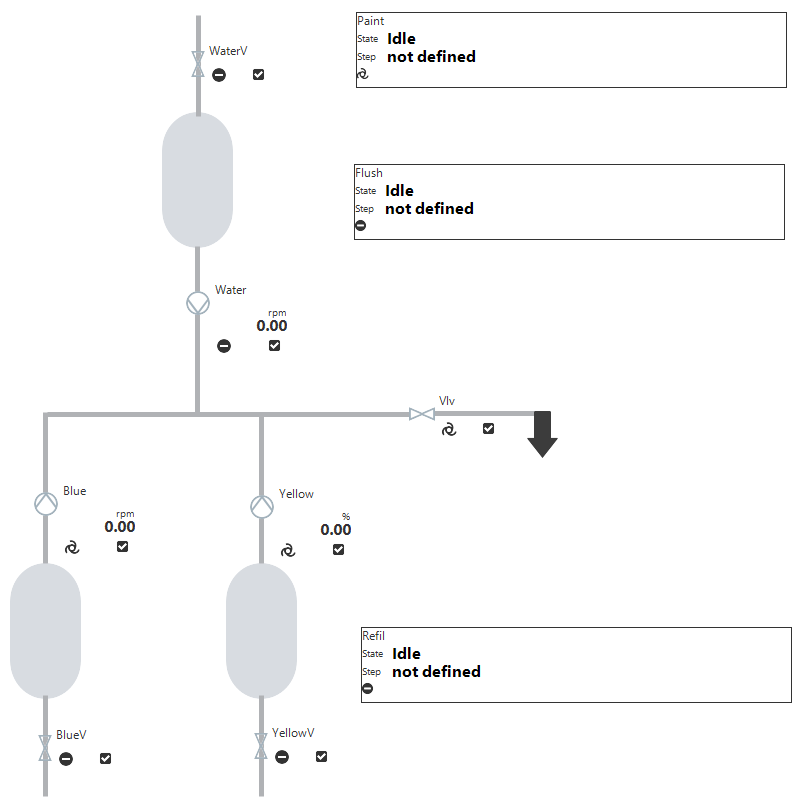}
    \caption{
    User interface of the Painting-Module with the control panels for the MTP services: Paint, Flush and Refil on the right hand side.
    }
    \label{fig:PMUI}
\end{figure}

One of the stations of the BaSys demonstrator is the painting module, a virtual \ac{PEA} automated with the MTP standard. It features two reservoirs of paint, one with blue and one with yellow ink, both equipped with pumps and process valves. It offers three skills as MTP services: the paint service can color tins situated inside the module either in blue or in yellow by spraying. Additionally it offers a refill service for the reservoirs, and a self-cleaning service using water from a third tank. The user interface of the module is shown in Fig. \ref{fig:PMUI}.

\subsection{Capability \ac{SM} realization}
\label{cap_sm_ralization}
\begin{figure}[htbp]
    \centering
    \includegraphics[width=\linewidth]{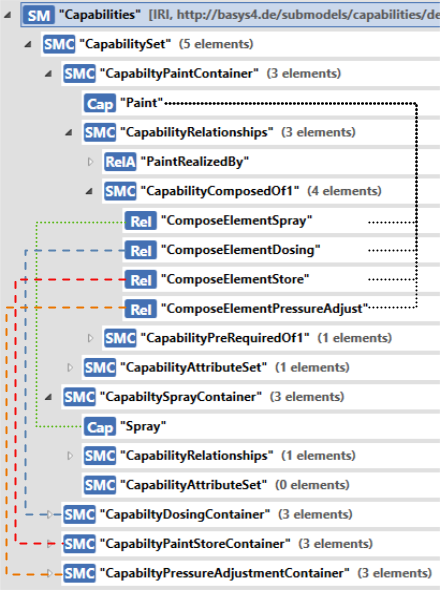}
    \caption{
    Realization of the painting module \ac{AAS} \ac{SM} with the \emph{composedOf} relations visualized, based on \ref{cap_tm_implications}
    }
    \label{fig:cap_sm_impl}
\end{figure}

To satisfy the reference route from ontologies to skill via the capability \ac{SM} as shown in Fig. \ref{fig:cap_sm_refs}, we realized a capability \ac{SM} for the painting module based on the equally named painting module example given in section \ref{cap_tm_implications}. The realization is shown in Fig. \ref{fig:cap_sm_impl}.

As already demonstrated in the example, the capability \emph{paint} is a composition of the capabilities \emph{spray}, \emph{dosing}, \emph{store} and \emph{pressure adjust}. As per the structure introduced in Section \ref{cap_sm}, every capability has its own \ac{SMC} with the contained capability \ac{SM} meta model element and its underlying structure. With this, every capability can reference an abstract capability defined within an ontology. Furthermore, by expressing the whole structure, every capability can be expressed as a composition of other capabilities and allow for a realization reference to a matching skill.

\subsection{Limitations \& Discussion}




In order to capitalize on a semantic capability description of production modules, it is key to have a common understanding of how to describe the capabilities. Our approach to use the established TGL 25000 standard for \emph{basic process operations} and relate it to \emph{process functions} is a good starting point. However, it needs to be completed and standardized to leverage its full potential in the engineering and operations phases. Especially the \emph{process functions} are lacking an accepted and complete taxonomy as of now.
For common applications, a universal ontology is generally possible, while it presumably remains necessary for vendors of specialized equipment to build upon this, and to extend the existing ontology with particular capabilities and their relations. It remains to be investigated further if all required extensions can be constructed from the elementary \emph{process functions} of the universal ontology. 

The capability description is not a static document. While the original capability submodel of a \ac{PEA} is supposed to be provided by the module vendor, it is possible that a \ac{PEA} is used in a different  function compared to what is has been designed for. In this case, the capability description might also be extended by the system integrator or plant owner. 

A limitation of the approach is the fact that validating the capabilities for a given requirement is not enough to ensure that the given module is capable of carrying out the actual task. Resources might have multiple constraints as in under which conditions capabilities can be offered up to which extend. In order to fully specify the scope of provided capabilities, as also briefly discussed in Section \ref{cap_sm}, a large number of attributes might be necessary leading to very complex descriptions which are difficult to standardize.  This could already be the case for simple modules: a tempering unit, for example, might have numerous attributes like \emph{heating power}, \emph{cooling power}, \emph{temperature range}, \emph{thermal stability}, \emph{reaction time},  \emph{backup power} and many more. For the purpose of automatic selection of modules, it is important to compare these attributes among different \acp{PEA}. This can be achieved with a standardized sets of attributes. The AAS provides a suitable framework to implement this using e.g. the technical data sheet submodel. We call a validation of capabilities including all their attributes a feasibility check. It gives more confidence in a suitable selection of modules, however, even with a feasibility check further validation might be needed. This could be done for example with virtual commissioning using simulation models of the given resources.

\section{Summary and Outlook}
In this paper, we have presented an approach to semantically describe the services offered by an MTP module. Therefore, an existing capability ontology has been extended by concepts that are specific to modules in the process industry. Additionally, the analogy of MTP services and \emph{skills} has been highlighted. This analogy was then used to realize the semantic annotation of MTP services based on an existing capability submodel for the \ac{AAS}.

The proposed approach allows treating MTP services like any other form of skills. This has the benefit that enhanced tooling\,--\,e.g. for automatically selecting modules' services based on capability requirements\,--\,can be implemented in a generic way that is agnostic of any MTP or process industry specifics. This is particularly interesting for hybrid industries, which combine process automation and factory automation technologies, but it contributes also to a more homogeneous automation landscape in general.  

If the selection of modules can be fully automated,
such tooling can be used for advanced \emph{Plug \& Produce} scenarios where modules' services are selected at runtime of a plant without extended engineering or planning phases. This provides the possibility to optimize the usage of the available resources such as \acp{PEA} within a plant or even globally.   

The presented extensions of the capability submodel for the AAS that allow to describe the composition of capabilities can also be reused for applications within other domains like the factory automation and is thus another step towards unification between process and factory automation. Another interesting use case is the modelling of capability requirements, i.e. the capabilities a \ac{PEA} needs from external devices in order to be complete for a given application. A use case for this would be direct module-to-module communication where either modules share equipment or together fulfill a common task where a redundant communication path is needed.



\section*{Acknowledgment}
The authors were partially supported by German Federal Ministry of Education and Research in the scope of the project BaSys 4.2 (01IS16022).

\vspace{0pt}
\bibliographystyle{IEEEtran}
\bibliography{bibliography}

\begin{thebibliography}{10}
\providecommand{\url}[1]{#1}
\csname url@samestyle\endcsname
\providecommand{\newblock}{\relax}
\providecommand{\bibinfo}[2]{#2}
\providecommand{\BIBentrySTDinterwordspacing}{\spaceskip=0pt\relax}
\providecommand{\BIBentryALTinterwordstretchfactor}{4}
\providecommand{\BIBentryALTinterwordspacing}{\spaceskip=\fontdimen2\font plus
\BIBentryALTinterwordstretchfactor\fontdimen3\font minus
  \fontdimen4\font\relax}
\providecommand{\BIBforeignlanguage}[2]{{%
\expandafter\ifx\csname l@#1\endcsname\relax
\typeout{** WARNING: IEEEtran.bst: No hyphenation pattern has been}%
\typeout{** loaded for the language `#1'. Using the pattern for}%
\typeout{** the default language instead.}%
\else
\language=\csname l@#1\endcsname
\fi
#2}}
\providecommand{\BIBdecl}{\relax}
\BIBdecl

\bibitem{vdi2019mtp1}
V.-G.~M. und Automatisierungstechnik~(GMA), ``Vdi/vde/namur 2658 part 1:
  Automation engineering of modular systems in the process industry -- general
  concept and interfaces,'' 2019.

\bibitem{packml}
\BIBentryALTinterwordspacing
(2022) What is packml. [Online]. Available: \url{https://www.omac.org/packml}
\BIBentrySTDinterwordspacing

\bibitem{vdi2022modularplants}
C.~Bramsiepe \emph{et~al.}, ``{VDI-Handlungsempfehlung -- Modulare Anlagen --
  Paradigmenwechsel im Anlagenbau},'' 2022.

\bibitem{weser2020ontology}
M.~Weser, J.~Bock, S.~Schmitt, A.~Perzylo, and K.~Evers, ``An ontology-based
  metamodel for capability descriptions,'' in \emph{2020 25th IEEE
  International Conference on Emerging Technologies and Factory Automation
  (ETFA)}, vol.~1.\hskip 1em plus 0.5em minus 0.4em\relax IEEE, 2020, pp.
  1679--1686.

\bibitem{iec2018aml}
IEC, ``Iec 62714-1: Engineering data exchange format for use in industrial
  automation systems engineering - automation markup language - part 1:
  Architecture and general requirements,'' 2018.

\bibitem{vdi2020mtp3}
{VDI/VDE-Gesellschaft Mess- und Automatisierungstechnik (GMA)}, ``Vdi/vde/namur
  2658 part 3: Automation engineering of modular systems in the process
  industry -- library for data objects,'' 2019.

\bibitem{vdi2019mtp2}
V.-G.~M. und Automatisierungstechnik~(GMA), ``Vdi/vde/namur 2658 part 1:
  Automation engineering of modular systems in the process industry --
  modelling of humand-machine-interfaces,'' 2019.

\bibitem{vdi2020mtp4}
{VDI/VDE-Gesellschaft Mess- und Automatisierungstechnik (GMA)}, ``Vdi/vde/namur
  2658 part 4: Automation engineering of modular systems in the process
  industry -- modelling of module services,'' 2020.

\bibitem{PlattI4_0:Glossary:D}
{Plattform Industrie 4.0}, ``Glossary,''
  \url{https://www.plattform-i40.de/SiteGlobals/IP/Forms/Listen/Glossar/DE/Glossar\_Formular.html?
  resourceId=1022602\&input\_=1020842\&pageLocale=de\&titlePrefix=D\#
  form-1022602}, (Accessed on 23/02/2022).

\bibitem{aasp1v3rc1}
M.~Sauer \emph{et~al.}, ``{Details of the Asset Administration Shell - Part 1 -
  The exchange of information between partners in the value chain of Industrie
  4.0 (Version 3.0RC01)},'' 11 2020.

\bibitem{gruener2021mtpaas}
S.~Gr{\"u}ner, M.~Hoernicke, A.~Kehl, C.~Barth, M.~Freund, M.~Hoffmeister, and
  T.~Klausmann, \emph{Integration of Module Type Package and Industry 4.0 Asset
  Administration Shell}.\hskip 1em plus 0.5em minus 0.4em\relax VDI Verlag,
  Düsseldorf, 01 2021, pp. 421--432.

\bibitem{PPR}
K.~Feldmann, T.~Schmuck, M.~Brossog, and J.~Dreyer, ``Beschreibungsmodell zur
  planung von produktionssystemen. entwicklung eines beschreibungsmodells
  f{\"u}r produkte, prozesse und ressourcen zur rechnergest{\"u}tzten planung
  produktionstechnischer systeme,'' \emph{wt Werkstattstechnik}, vol.~98,
  no.~3, 2008.

\bibitem{jwp_cap_skill}
A.~Bayha, J.~Bock, B.~Boss, C.~Diedrich, and S.~Malakuti, ``{Describing
  Capabilities of Industrie 4.0 Components - Joint White Paper between
  Plattform Industrie 4.0, VDI GMA 7.20, BaSys 4.2},'' 12 2020.

\bibitem{aas2020details}
S.~Bader, E.~Barnstedt, H.~Bedenbender, M.~Billman, B.~Boss, and A.~Braunmandl,
  ``Details of the asset administration shell part 1-the exchange of
  information between partners in the value chain of industrie 4.0 (version 3.0
  rc01),'' \emph{German Electrical and Electronics Manufacturers Association,
  Frankfurt am Main, Germany}, 2020.

\bibitem{functionalCapabilitySkill}
W.~Motsch, K.~Dorofeev, K.~Gerber, S.~Knoch, A.~David, and M.~Ruskowski,
  ``Concept for modeling and usage of functionally described capabilities and
  skills,'' in \emph{2021 26th IEEE International Conference on Emerging
  Technologies and Factory Automation (ETFA )}, 2021, pp. 1--8.

\bibitem{tgl25000}
{Amt für Standardisierung, Messwesen und Warenprüfung der Deutschen
  Demokratischen Republik (DDR)}, ``Tgl 25000: Verfahrenstechnik
  grundoperationen,'' 1974.

\end{thebibliography}

\end{document}